\documentclass[amsmath,amssymb,amsfonts,aps,prl,twocolumn,superscriptaddress,footinbib]{revtex4-2}

\usepackage{graphicx}
\usepackage{dcolumn}
\usepackage{bm}
\usepackage[normalem]{ulem}
\usepackage{xcolor}
\usepackage{comment}
\usepackage{amsmath}
\usepackage{siunitx}
\usepackage{dsfont}
\usepackage{float}
\usepackage{placeins}
\usepackage[shortlabels]{enumitem}
\usepackage{xr}
\usepackage{booktabs}

\DeclareSIUnit\angstrom{\text{Å}}

\begin{document}

\title{Spin-valve effect through electrostatic coupling}
\date{\today}

\author{J. Phillips }
\affiliation{International Iberian Nanotechnology Laboratory (INL), Av. Mestre Jos\'e Veiga, 4715-330 Braga, Portugal}
\email{jan.phillips@inl.int}

\author{J. Fern\'andez-Rossier}
\altaffiliation[On permanent leave from ]{Departamento de F\'isica, Universidad de Alicante, 03690 San Vicente del Raspeig, Spain.}
\affiliation{International Iberian Nanotechnology Laboratory (INL), Av. Mestre Jos\'e Veiga, 4715-330 Braga, Portugal}
\email{joaquin.fernandez-rossier@inl.int}

\begin{abstract}
We propose a spin-valve effect in which an insulating magnetic van der Waals multilayer controls the conductance of an adjacent non-magnetic conductor by electrostatic coupling, without transport-electrons entering the magnetic layers. The coupling is induced by the variations of the electronic structure when the relative orientation of the magnetization of adjacent monolayers changes from parallel to antiparallel, modifying both the dielectric constant and band alignment.
We consider two types of coupling and device concepts: charge transfer to a graphene monolayer and capacitive coupling in a single-electron transistor. We use density functional calculations to assess the magnitude of the effect in various materials, including CrI$_3$, CrBr$_3$, CrCl$_3$ and NiBr$_2$. We predict magnetoresistance ratios in the range of 1.14-1.20 for the charge-transfer device and up to $10^3$ in the single-electron transistor.

\end{abstract}

\maketitle

%\section{Introduction}
{\em Introduction ---}
Spin valves are layered devices comprising at least two magnetic layers whose relative magnetization orientation controls their electrical conductance \cite{dieny91b}. A small magnetic field can switch their relative orientation, producing a large change in electrical resistance. The spin-valve concept emerged from the discovery of giant magnetoresistance (GMR) in artificial multilayers of alternating ferromagnetic and nonmagnetic metals \cite{Bai88}. Spin-valve functionality is also realized in magnetic tunnel junctions (MTJs), where two ferromagnetic electrodes are separated by an ultrathin insulating barrier, giving rise to tunnel magnetoresistance (TMR) \cite{julliere1975tunneling, parkin99,parkin04}. Spin valves are central to spintronics, underpinning technologies ranging from highly sensitive magnetic-field sensors to nonvolatile magnetic memories \cite{parkin2003magnetically,freitas07}.

\begin{figure}[h!]
    \centering
    \includegraphics[width=\linewidth]{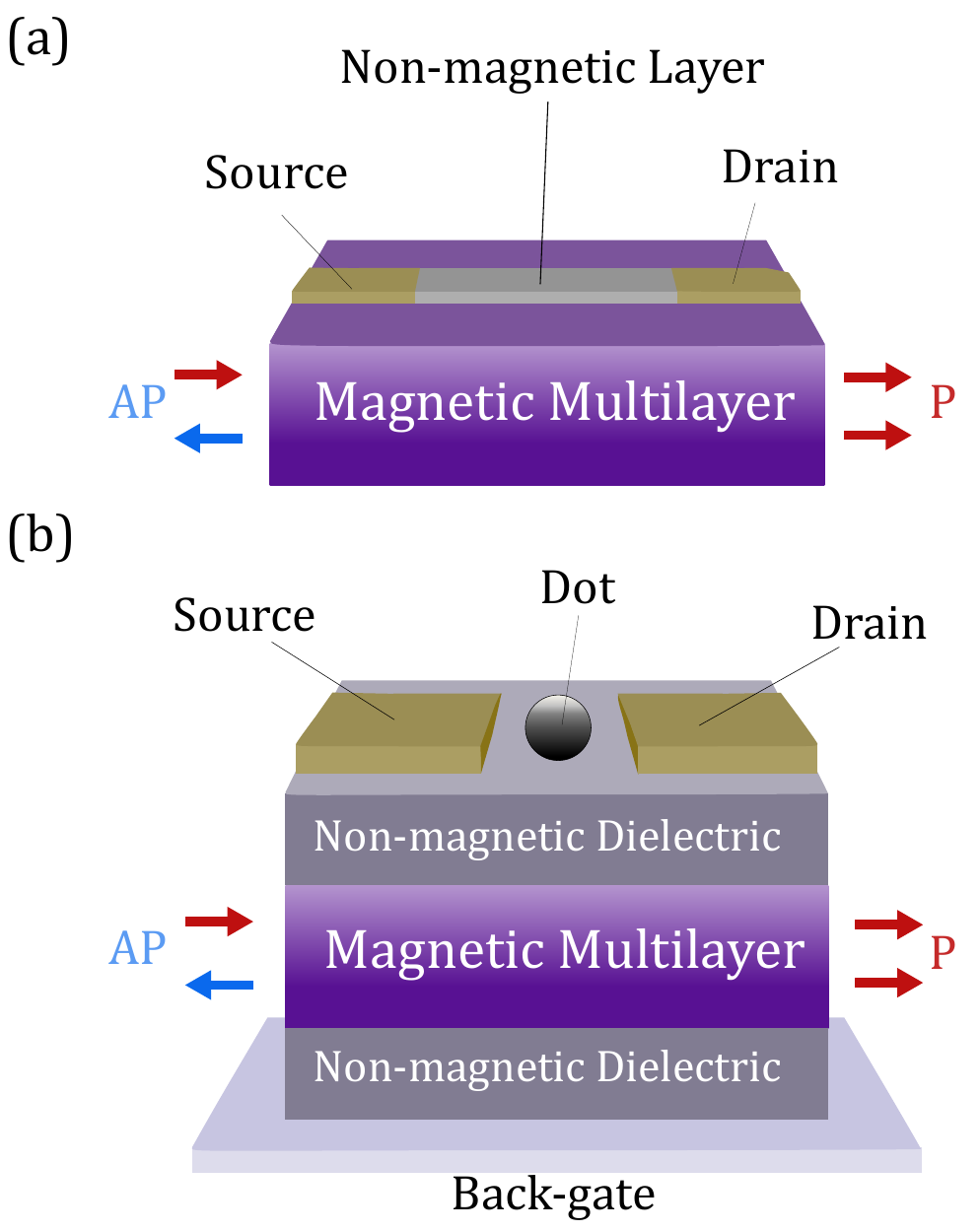}
    \caption{Schematic picture of the spin-valve devices: (a) Charge transfer to an adjacent conductor and (b) capacitive coupling to a SET (or quantum dot). The SET device consists of a metallic island, contacted by source and drain electrodes, which lies on the hBN-encapsulated magnetic multilayer. Both devices are based on the possibility of switching between a P and AP state of the insulating magnetic multilayer.}
    \label{fig:device}
\end{figure}

%\section{Device}
Conventional spin-valves rely predominantly on spin-dependent scattering\cite{dieny91} and spin-dependent tunnelling\cite{parkin99,parkin04} through the magnetic layers themselves: the transport electrons must traverse or scatter within the magnetic stack.
Here, we propose a spin-valve mechanism that is {\em fundamentally different } from previous mechanisms and platforms in that the transport electrons never enter the magnetic layers \footnote{In our calculations we see that the charge transferred to the magnetic layers occupies a flat, low-mobility conduction band in such a way that the conduction takes place in the non-magnetic layer alone.}.
Instead, the magnetic multilayer acts purely electrostatically on an adjacent non-magnetic conductor. This is also different from the van der Waals spin valve~\cite{cardoso18}, which relied on magnetic proximity effect.

We propose two types of transduction mechanism for these {\em  electrostatic-coupling spin valves}: charge transfer to an adjacent conductor (graphene) and capacitive coupling to a  single-electron transistor (SET), shown in Fig.~\ref{fig:device}(a) and (b) respectively. Graphene converts the magnetically-modulated work function due to the change in band-gaps into a continuous change in carrier density and therefore in sheet conductance. We expect the mechanism to apply more broadly to any conductor whose electrical response is sensitive to the local electrostatic potential set by a neighboring interface. In the SET, capacitive coupling occurs through the dependence of the dielectric function of the insulating magnetic layer on its magnetic state. 

Charge transfer between a magnetic insulator and an adjacent non-magnetic conductor is itself an established phenomenon. Graphene placed next to a chromium halide is known to become doped by interfacial charge transfer and has been shown experimentally to change discontinuously when a CrI$_3$ bilayer's magnetic order is switched between parallel and antiparallel states~\cite{tseng2022gate}. Related magnetically-tunable charge transfer has also been reported in graphene/$\beta$-RuCl$_3$~\cite{razpopov2024} and graphene/CrSBr~\cite{kundu2026crsbr}. Here we identify the microscopic origin of this effect. 

\begin{figure}
    \centering
    \includegraphics[width=\linewidth]{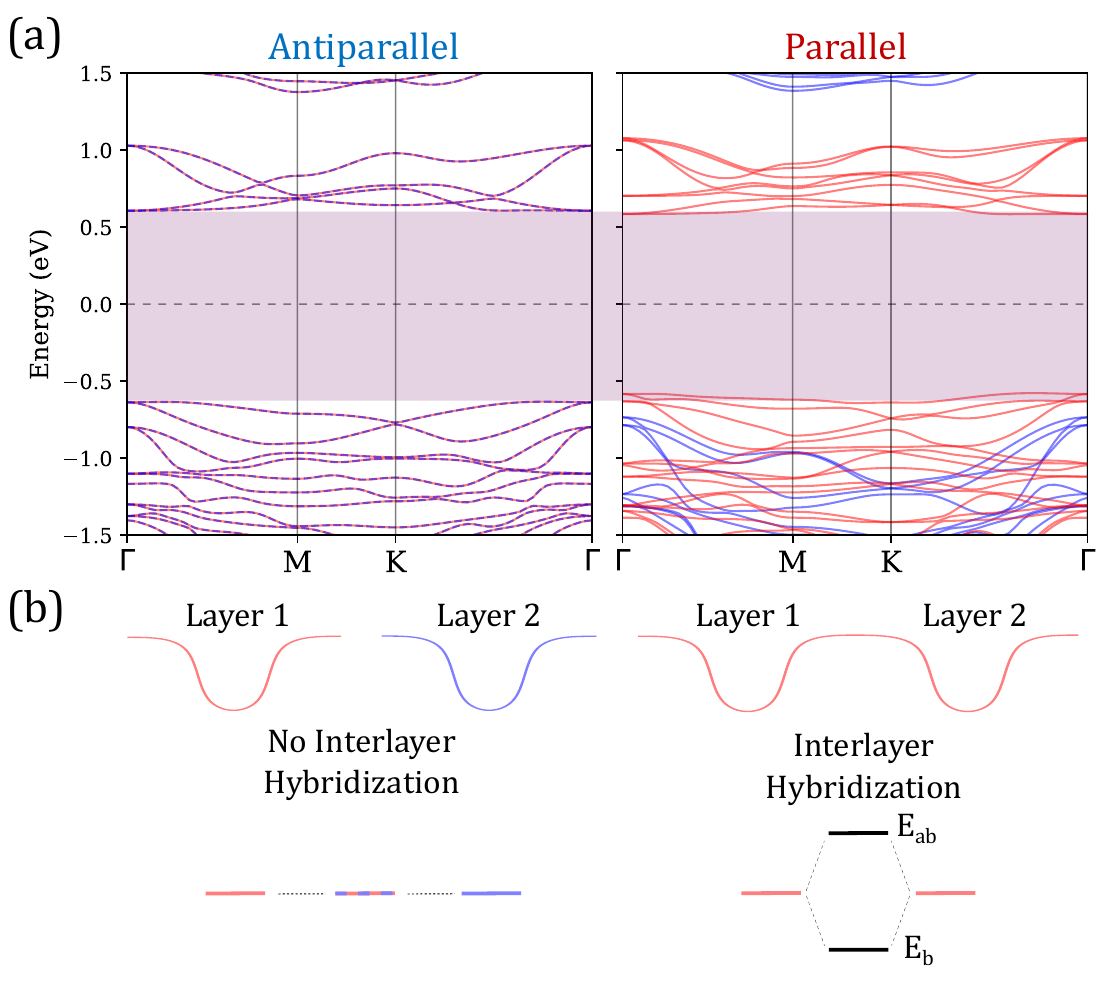}
    \caption{(a) DFT Band structures of an insulating bilayer for an AP (left) and P (right) magnetic order of the spins between layers. Blue (red) lines represent the minority (majority) spin-channel.  The shaded area represents the band gap for the antiparallel configuration, making it obvious by comparison that the P band gap decreases. The specific material used for this representation was bilayer CrI$_3$. (b) Schematic of the two insulating layers, for the AP (left) and P (right) magnetic orders. Layers are represented as potential wells, showing how the energy levels remain degenerate in the antiparallel situation, while the energy levels hybridize forming bonding and anti-bonding states in the parallel case.} 
    \label{fig:band_gap}
\end{figure}

{\em Impact of the magnetic state on the electronic properties of the magnetic multilayer ---} 
We now discuss how the relative orientation of the magnetization in  van der Waals multilayers can affect the value of their band-gap, their work function and the dielectric constant. In a wide class of materials, including the Cr trihalides, the atomic orbitals forming the valence and conduction bands are fully spin polarized. In the P state, interlayer hopping hybridizes the bands of the two monolayers and widens them (Fig.~\ref{fig:band_gap}(b)); in the AP state this hybridization is forbidden, since the corresponding states of the two layers have opposite spin, and the bilayer bands are two copies of the monolayer ones. The gap is therefore smaller in the P state, as confirmed by DFT for the bilayer (Fig.~\ref{fig:band_gap}(a)) and the bulk (Table~\ref{tab:gaps_eps}). Since this hybridization redistributes charge across the van der Waals gap, it also changes the surface dipole of the bilayer, and with it the work function (Supplementary Material). 

The dielectric constant is instead governed by the optical gap. In perturbation theory, the out-of-plane susceptibility follows from virtual interband transitions,
\begin{equation}
    \chi_{zz} \propto \sum_n \frac{|\langle 0|z|n\rangle|^2 }{E_n - E_0},
    \label{eq:polarizability}
\end{equation}
where $|0\rangle$ is the electronic ground state and $|n\rangle$ are excited states reached by the \emph{spin-conserving} promotion of an electron from the valence to the conduction band. Since $\varepsilon=1+\chi$, $\varepsilon$ increases as the lowest such transition energy decreases\footnote{More elaborate models such as the Penn model~\cite{Penn1962} give $\varepsilon-1\propto1/E_g^2$. They describe the electronic contribution $\varepsilon^\infty$, which is the quantity computed here.}. In the Cr trihalides both band edges are majority-spin (Fig.~\ref{fig:band_gap}(a)), so the fundamental gap is dipole-allowed and coincides with the optical gap. In NiBr$_2$ the conduction band minimum is minority-spin, so in the P state the fundamental gap is a spin-flip transition, dipole-forbidden to leading order. As a result, while the fundamental gap narrows from AP to P in all four materials, $\varepsilon^\infty_{zz}$ increases in the Cr trihalides but decreases in NiBr$_2$ (Table~\ref{tab:gaps_eps}). The work function, which depends on band-edge positions rather than on optical matrix elements, is instead controlled by the fundamental gap\footnote{As shown in the Supplementary Material, $\delta\Phi\approx e\,\delta\Delta V_{\rm surf}+(\tfrac12-f)\,\delta E_g$, where $f$ is the fraction of the gap change carried by the conduction band.}. The magnetic-state dependences of $\Phi$ and $\varepsilon$ are thus governed by distinct gaps.

\begin{table}[h]
\centering
\begin{tabular}{ccccccccc}
\toprule
 & \multicolumn{2}{c}{CrI$_3$} & \multicolumn{2}{c}{CrBr$_3$} & \multicolumn{2}{c}{CrCl$_3$} & \multicolumn{2}{c}{NiBr$_2$} \\
\cmidrule(lr){2-3} \cmidrule(lr){4-5} \cmidrule(lr){6-7} \cmidrule(lr){8-9}
 & AP & P & AP & P & AP & P & AP & P \\
\midrule
E$^F_g$ (eV) & 1.18 & 1.04 & 1.52 & 1.38 & 1.80 & 1.68 & 0.89 & 0.62 \\
E$^O_g$ (eV)     & 1.18 & 1.04 & 1.52 & 1.38 & 1.80 & 1.68 & 0.89 & 1.05 \\
$\varepsilon^\infty_{zz}$  & 5.76 & 6.26 & 4.05 & 4.23 & 3.43 & 3.51 & 6.22 & 4.97 \\
$\delta\Phi$ (meV) & \multicolumn{2}{c}{20.00} & \multicolumn{2}{c}{9.28} & \multicolumn{2}{c}{9.70} & \multicolumn{2}{c}{33.31} \\
\bottomrule
\end{tabular}
\caption{DFT-derived fundamental gap ($E_g^F$), optical gap ($E_g^O$), and electronic out-of-plane dielectric constant $\varepsilon^\infty_{zz}$ for the parallel (P) and antiparallel (AP) magnetic configurations of each bulk material. Variation of the work function ($\delta\Phi=\Phi^P-\Phi^{AP}$) when changing the magnetic state of the system, computed from DFT for the bilayers with the Fermi level placed at midgap. Positive (negative) values mean that the P (AP) phase has a larger work function $\Phi$.}
\label{tab:gaps_eps}
\end{table}

\begin{figure}
    \centering
    \includegraphics[width=\linewidth]{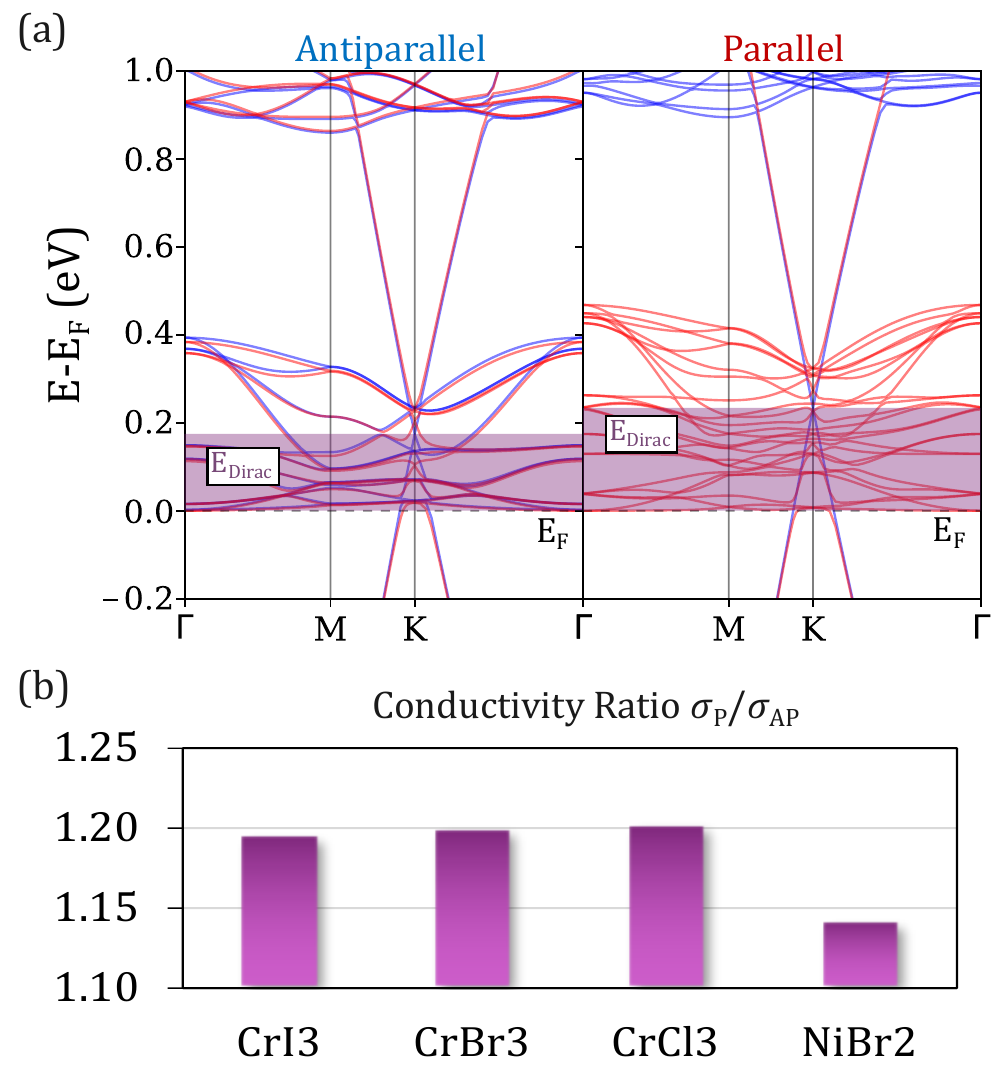}
    \caption{(a) Bandstructure for a hetero-structure consisting of graphene on top of a CrI$_3$ %magnetic insulating 
    bilayer. Blue (red) lines represent the minority (majority) spin-channel. The shaded area is a guide to the eye to show where the graphene Dirac cone falls within the insulating bilayer bands. (b) Graphene conductivity ratio $\sigma_P/\sigma_{AP}$. Results obtained  solving Eqs. \ref{eq:graph_capacitor} and \ref{eq:cond_change}.}
    \label{fig:graphene}
\end{figure}

{\em Charge-transfer spin valve ---}
Here we demonstrate how the different magnetic state affects the charge transfer from the non-magnetic layer to the magnetic bilayer substrate. For this, we required a heterostructure unit cell (details in the Supplementary Material) to perform the DFT calculations. The work function sets the interfacial charge transfer~\cite{mott1938note} and the doping of the conductor therefore depends on the magnetic state, as do the metal-induced gap states~\cite{heine1965theory, tersoff1984theory, tersoff1985schottky, tejedor1977metal}. Fig. \ref{fig:graphene} shows the bandstructure of the combined system when switching from the two magnetic configurations: AP and P. It becomes obvious that the Dirac cone of the chosen non-magnetic layer (graphene) is within what used to be the conduction bands of the chosen magnetic bilayer (CrI$_3$) in both situations. The Fermi level is now below the characteristic Dirac cone, meaning that charge is transferred in this situation from the non-magnetic layer to the substrate. Importantly, the Fermi level is lower down from the Dirac cone in the P magnetic configuration, translating into a higher charge transfer compared to that of the AP configuration. 

We now estimate how much graphene's conductance $\sigma$ changes when switching between the two magnetic states, P and AP, of the van der Waals bilayers. To do so, we assume $\delta\sigma\propto \delta n$, where $\delta n$ is the density of carriers transferred across the interface, which we infer from the work-function difference~\cite{wigner1935theory, cardoso2023strong} 
\begin{equation}
\delta n= \varepsilon_0\frac{\Phi_{bilayer} - \Phi_{graphene}}{e^2\,d}
\equiv\varepsilon_0\frac{\Delta \Phi}{e^2\,d}
%\Phi_2 - \Phi_1 = \frac{e^2\,\delta n}{\varepsilon_0}\,d,
\label{eq:graph_capacitor}
\end{equation}

Eq. \ref{eq:graph_capacitor} basically treats the interface between graphene and the van der Waals insulator as a parallel plate capacitor\footnote{Since $\Phi_{\rm bilayer} - \Phi_{\rm graphene}$ is extracted from the self-consistent heterostructure, graphene's quantum capacitance~\cite{giovannetti2008doping} is already included}, separated by a distance $d$, with a voltage drop given by the work-function difference of the two materials. We obtain  $\Delta \Phi$  from DFT calculations of the heterostructures, using the planar-averaged electrostatic potential along $z$, with $E_{\rm vac}$ taken from its plateau in the vacuum region. We include dipole corrections~\cite{neugebauer1992adsorbate, bengtsson1999dipole}, as is standard when work functions are extracted from asymmetric slabs~\cite{giovannetti2008doping, khomyakov2009first, ding2017prediction}, so that $\Delta\Phi$ corresponds to the voltage drop across the heterostructure. We carry out this calculation in two magnetic states $P$, $AP$ and obtain different values of $\Delta \Phi$. 

Using the  linear relation between conductance and carrier density\cite{novoselov2005graphene}, together with Eq. \ref{eq:graph_capacitor} we estimate the ratio between the P and AP associated graphene conductances as:

\begin{equation}
    \frac{\sigma_P}{\sigma_{AP}}= \frac{\Delta \Phi_{P}}{\Delta \Phi_{AP}}
    \label{eq:cond_change}
\end{equation}

The resulting conductivity ratios are shown in Fig.~\ref{fig:graphene}(b) (values of $\Delta\Phi$ in Table~SIII of the Supplementary Material), taking the lower conductivity state as a reference. Our calculations yield ratios of $\sigma$ in the range of 1.14-1.20, which serves as an estimate of the efficiency of the proposed charge-transfer spin valve concept in Fig.~\ref{fig:device}(a).

\begin{figure}
    \centering
    \includegraphics[width=\linewidth]{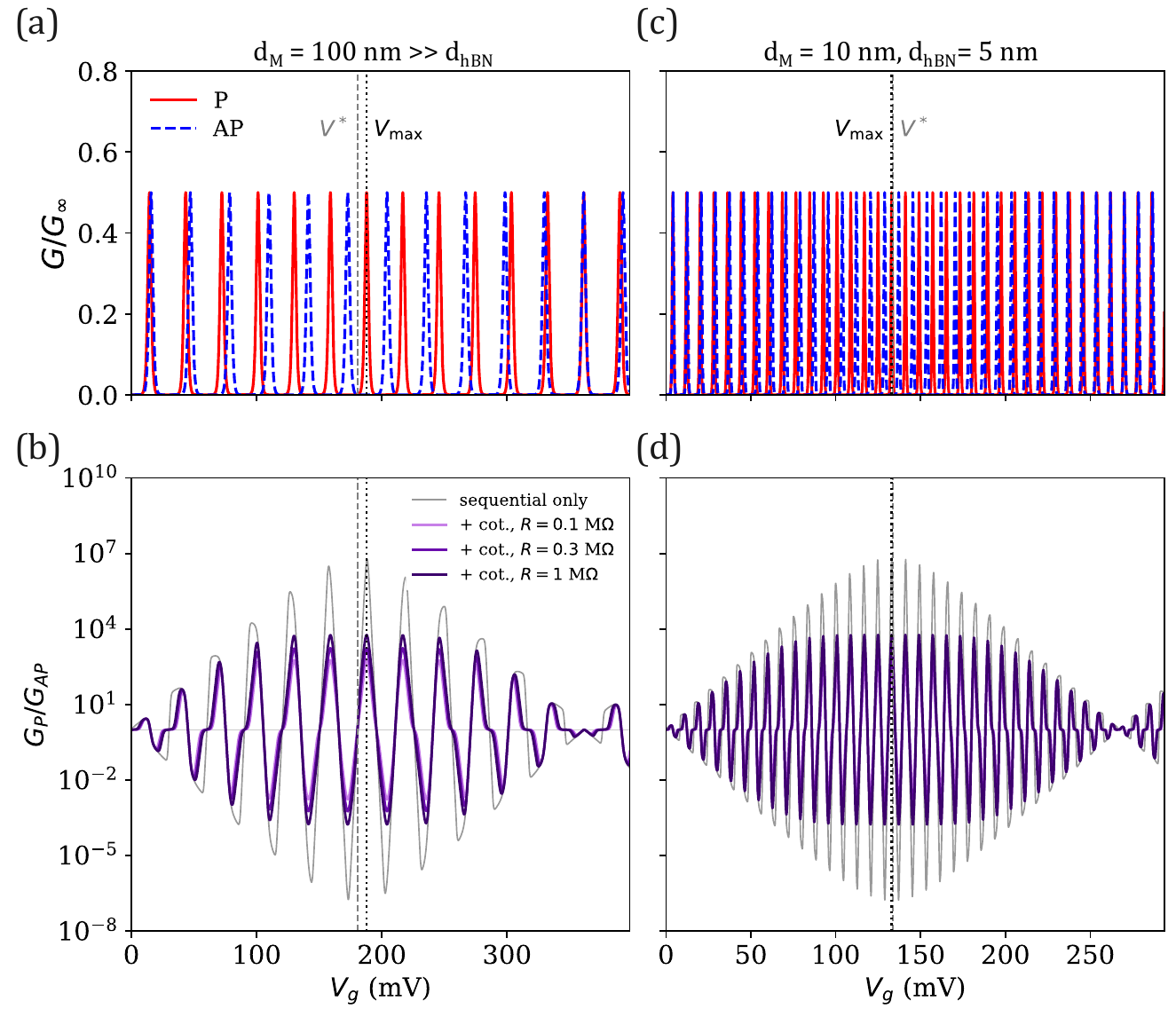}
    \caption{SET realization of the spin-valve effect through electrostatic
    coupling, with CrI$_3$ as the magnetic multilayer. Top row: normalized
    conductance $G/G_\infty$ as a function of gate voltage $V_g$ in the P (red)
    and AP (blue) states. Bottom row: conductance ratio $G_\mathrm{P}/G_\mathrm{AP}$
    for the same configurations. (a),(b) Thick-multilayer limit,
    $d_M \gg d_\mathrm{hBN}$, where the gate capacitance is set by the
    multilayer dielectric constant alone; (c),(d) finite stack with
    $d_M = 10$~nm and $d_\mathrm{hBN} = 5$~nm. The dashed line marks the P conductance
    peak closest to $V^*$, where the P and AP peak sequences are out of phase.}
    \label{fig:SET}
\end{figure}

{\em Capacitively controlled spin valve ---} We continue to present our second device concept, based on the capacitive coupling in a SET. As we show below (Fig.~\ref{fig:SET}), this provides a substantial {\em modulation} of the conductance between the two magnetic states of the van der Waals magnet. We consider a dot where the dielectric layer separating back-gate from the source-dot-drain layer contains a magnetic multilayer (Fig.~\ref{fig:device}(b)), which is encapsulated by few-nanometer hBN layers~\cite{cadiz2017excitonic, frisenda2018recent, bie2017mote2}. Here the multilayer exchanges no charge with source-dot-drain layer, where transport takes place. Unlike the magneto-Coulomb effect~\cite{ono1997enhanced, wunderlich2006coulomb, seneor2007nanospintronics}, where the island or the electrodes are ferromagnetic conductors, here the magnetic material is an insulator integrated in the gate dielectric.

In order to estimate the spin-valve effect in this setup, we use the orthodox Coulomb-blockade theory~\cite{beenakker1991theory, likharev1999single} in the metallic limit $\Delta E\ll k_BT\ll E_C$, with $\Delta E$ the single-particle level spacing, $T$ the temperature and $E_C=e^2/2C_T$ the charging energy, where $C_T$ is the total capacitance of the island. In this regime, the linear conductance $G$ of the SET follows from the orthodox rate equations~\cite{beenakker1991theory}:

\begin{equation}
  \frac{G}{G_\infty}=\sum_N P_{\rm eq}(N)\,
  F\!\left(\frac{U(N)-U(N-1)}{k_BT}\right),
  \label{eq:G}
\end{equation}

where $F(x)=x/(1-e^{-x})$ and $N$ is the number of excess electrons on the island, $U(N)=E_C(N-n_g)^2$ its electrostatic energy, $n_g=(C_gV_g+q_0)/e$ the charge induced by the gate voltage $V_g$, and $P_{\rm eq}(N)\propto e^{-U(N)/k_BT}$ the equilibrium probability of state $N$. Here $G_\infty=(R_l+R_r)^{-1}$, with $R_l$ and $R_r$ the resistances of the two tunnel junctions. 

The magnetic state of the multilayer affects transport through the modification of the dielectric constant, that affects the gate capacitance $C_g$ that controls the charging curves of the SET. The gate stack forms capacitors in series over the island area $A$, 
\begin{equation}
C_g=\varepsilon_0A/\sum_i(d_i/\varepsilon_i)
\label{eq:Cg}
\end{equation}
where $d_i$ and $\varepsilon_i$ are the thickness and out-of-plane permittivity of each layer (hBN and magnetic multilayer (M)). Switching the magnetic state changes $C_g$ through $\varepsilon_M$, while all other capacitances and the offset charge $q_0$ remain unchanged (see End Matter).

Conductance peaks occur when the back-gate voltage is tuned to obtain  $U(N-1)=U(N)$:
\begin{equation}
    V_g^{(N)}=[e(N-\tfrac12)-q_0]/C_g
    \label{peaks}
\end{equation}

%%%%%%%%%%%%%%%%%%%%%
Hence, the $G(V_g)$ curve describes a comb of peaks, spaced by $\Delta V_g= \frac{e}{C_g}$ .  Since $P$ and $AP$ magnetic states have a different $C_g$, on account of Eq.~\ref{eq:Cg}, the $G(V_g)$ combs have {\em different }  
periods: $\Delta V_P < \Delta V_{AP} $ for the halides and reversed for NiBr$_2$.  
The combs start in phase at $V_g=0$ and drift apart as $V_g$ increases, as shown in Fig.~\ref{fig:SET}(a).   
 
It is apparent that the variation of conductance as the magnetic state changes between $P$ and $AP$ states  depends strongly on the gate voltage. This is most clearly seen in Fig.~\ref{fig:SET}(b). The two combs reach full antiphase at

\begin{equation}
  V^*=\frac{e}{2|C_g^P-C_g^{AP}|}%\simeq\frac{e\,d_M}{2\varepsilon_0A\,|\Delta\varepsilon_M|},
  \label{eq:V_anti}
\end{equation}

that provides a good estimate for the {\em maximal sensitivity} voltage, $V_{\rm max}$, defined as the voltage for which the  conductance difference between the P and AP states is maximal. $V_{\rm max}$ is the voltage closest to $V^*$ for which one of the magnetic states features a peak and the other a valley.

The variation of the capacitance in Eq. \ref{eq:V_anti} is driven by  the variation of the dielectric constant, $\Delta\varepsilon_M=\varepsilon_M^P-\varepsilon_M^{AP}$. When the width of the multilayer is much larger than the width of the non-magnetic layers on the gate stack,  $d_M/\varepsilon_M\gg2d_{\rm hBN}/\varepsilon_{\rm hBN}$, we can approximate  $\frac{1}{|C_g^P-C_g^{AP}|}\simeq\frac{d_M}{\varepsilon_0A\,|\Delta\varepsilon_M|}$.   
  %$V^*$ is independent of $C_T$ and $T$. 
 For the parameters of the End Matter, $V^*\approx0.18$, $0.50$, $1.13$ and $0.07$~V for CrI$_3$, CrBr$_3$, CrCl$_3$ and NiBr$_2$ (parallel-plate estimates, see End Matter)\footnote{ NiBr$_2$, whose dielectric constant decreases from AP to P, displaces its peaks in the opposite direction}.

We now discuss the magnitude of the conductance ratio, $G_P/G_{AP}$, as  $V_g$ is varied. If the valley conductance were limited to the single-electron transport, peak-to-valley conductance ratios predicted by the orthodox theory are in the range of $10^6$.  Experimentally\cite{saitoh04,saitoh05}, most   peak-to-valley ratios are below 10$^3$, because of other contributions to the valley conductance,  such as inelastic cotunneling, but higher ratios have been reported at low temperatures\cite{park05}. Inelastic cotunneling contribution scales as\cite{averin1990,turek2002cotunneling}
$G_{\rm cot}\simeq \frac{2}{3}\frac{R_K}{R^2}\left(\frac{k_B T}{E_C}\right)^2$, where $R_K=\frac{h}{e^2}$ and $R$ is the  electrode-dot resistance.

In Fig.~\ref{fig:SET}(b) we plot $G_P/G_{AP}$ for various values of the junction resistances of $0.1$--$1$~M$\Omega$~\cite{likharev1999single} which gives $G_P/G_{AP}\sim6\times10^2$--$6\times10^3$. Hence,  the figure of merit of capacitively-coupled  spin-valves is dramatically larger than the ones predicted for the charge-transfer  spin valve. A significant change is already reached once the peaks are displaced by about one peak width ($G_P/G_{AP}\sim\!8$, rising to $\sim\!10^2$ at two), so the operating point can be chosen anywhere between a flank and full antiphase. The scenario when the multilayer thickness is comparable to the hBN thickness is represented in Fig.~\ref{fig:SET}(c),(d), where the hBN layers form 50$\%$ of the gate stack, and show how the period of the conductance peaks change for both magnetic states, while the maximum accessible conductance ratios remain unchanged.

Magnetic-field control of the proposed devices can be realized using a multilayer with an AP  ground state at zero field, which turns into a P state  upon application of a sufficiently strong magnetic field. This configuration occurs naturally in A-type antiferromagnets, comprising ferromagnetically ordered layers coupled antiferromagnetically. Examples include bilayer and antiferromagnetically coupled few-layer CrI$_3$ \cite{huang2017layer, Klein18}, CrCl$_3$ \cite{Kim19}, and CrSBr \cite{Wilson2021Interlayer}. Alternatively, engineered magnetic stacks could provide the required configuration. Conventional magnetic tunnel junctions offer precedents for controlling magnetic configurations through exchange bias and interlayer coupling: exchange bias stabilizes a reference magnetization \cite{parkin99}, while synthetic ferrimagnetic structures have enabled stable P and AP junction configurations at zero applied field \cite{Hayakawa08}.

Control of the magnetic configuration also offers prospects for information storage. Reversible electrical switching between interlayer antiferromagnetic and ferromagnetic states has been demonstrated in bilayer CrI$_3$ near its field-induced transition \cite{Jiang18El}. All-optical magnetization reversal has also been demonstrated in atomically thin CrI$_3$ \cite{Zhang22}. If electrical or optical stimuli can switch a multilayer between distinguishable P and AP states, these configurations could encode a classical bit, with the adjacent conductor providing electrical readout through the mechanisms discussed here. 

Although our calculations compare the collinear P and AP configurations, the field-induced evolution between these limits need not be abrupt. Depending on magnetic anisotropy and field orientation, intermediate configurations can arise through continuous canting of the layer magnetizations, as demonstrated in CrCl$_3$ under an out-of-plane field \cite{Kim19}.  In a suitably engineered multilayer\cite{freitas07}, a continuous  response would allow small magnetic-field variations to be converted into conductance changes, providing a route to magnetometry based on the electrostatic transduction mechanisms
proposed here.

{\em Conclusions ---}
In conclusion, we propose a new paradigm for spin-valve operation in which the relative magnetization of insulating magnetic layers controls a separate conducting channel through electrostatic coupling. The magnetic configuration modifies charge transfer or gate capacitance, converting changes in the multilayer’s electronic properties into a conductance response without requiring spin dependent interactions of the transport electrons. Our DFT calculations provide order of magnitude estimates of the magnitude of the spin valve effects for both type of coupling, charge-transfer in graphene and capacitive coupling in  single-electron transistors, respectively. In the latter, Coulomb blockade amplifies modest dielectric changes into conductance ratios reaching \(10^{2}\)–\(10^{3}\) under the conditions considered. 
These results show that large spin-valve effects can emerge from the amplification of a magnetic insulator’s electrostatic response by Coulomb blockade, without requiring spin-dependent transport in the conducting channel.

\begin{acknowledgments}
We acknowledge financial support from 
%1
 %2 
 SNF Sinergia (Grant Pimag,  CRSII5\_205987). We thank  RES for computational resources provided by Barcelona Supercomputing Centre in MareNostrum 5 through project FI-2025-3-0039.
 %3
J.F.-R. acknowledges financial funding from 
% 4
Generalitat Valenciana (Prometeo2021/017
and MFA/2022/045)
%5
and
MICIN-Spain (Grants No.  PID2022-141712NB-C22) %and PRTR-C1y.I1).
We acknowledge the use of OpenAI's ChatGPT and Anthropic's Claude as  research aids for theoretical discussions, 
and manuscript editing. We have carefully revised all the results. Accordingly, we take full responsibility for the presented material.

\end{acknowledgments}

\nocite{QE-2009, QE-2017, doi:10.1063/5.0005082, PBE1996, van_Setten_2018, Hamann2013, Grimme2010, Blochl1994}
\bibliography{bib}
\clearpage    

\section*{End Matter}
\appendix

\section*{SET model: details and parameters}
\label{app:SET}

\textit{Gate stack.}---We treat the stack as parallel-plate capacitors in series over the island area $A$:
\begin{equation}
  \frac{1}{C_g}=\frac{1}{\varepsilon_0A}\left(
  \frac{2d_{\rm hBN}}{\varepsilon_{\rm hBN}}+\frac{d_M}{\varepsilon_M}\right).
\end{equation}

Figs.~\ref{fig:SET}(a),(b) are computed in the multilayer-dominated limit, $d_M/\varepsilon_M\gg2d_{\rm hBN}/\varepsilon_{\rm hBN}$; setting $d_{\rm hBN}=0$ in Table~\ref{tab:SET_param} is a statement about the capacitance, not the absence of these layers. Figs.~\ref{fig:SET}(c),(d) use $d_M=10$~nm and $d_{\rm hBN}=5$~nm, with all other parameters unchanged. The encapsulating hBN layers do not depend on the magnetic state: they reduce both $C_g$ and its relative change upon switching, and therefore increase $V^*$ at fixed $d_M$. For $d_M=100$~nm and $\varepsilon_{{\rm hBN}, \perp}\approx3.4$~\cite{laturia2018dielectric}, hBN layers of $5$~nm increase $V^*$ by a factor $\approx1.4$, and $30$~nm layers by a factor $\approx4$. In Figs.~\ref{fig:SET}(c),(d), by contrast, the thinner multilayer raises $C_g$ and lowers $V^*$ to $\approx0.14$~V. Only the out-of-plane component of the permittivity enters, since the gate field is normal to the layers. We use the electronic value $\varepsilon^\infty_{zz}$ from DFPT. The quasistatic gate sweep also probes the ionic contribution, but since this depends only weakly on the relative orientation of the layer magnetizations, it largely cancels in $\Delta\varepsilon_M$ and mainly rescales $C_g$.

\textit{Charging energy.}---We write $C_T=C_g+C_{\rm rest}$, where $C_{\rm rest}=C_s+C_d+C_0$ collects the capacitances not involving the gate and is independent of the magnetic configuration. For the illustrative value $C_T^P=100$~aF (Table~\ref{tab:SET_param}), $C_T^{AP}=C_T^{P}-C_g^{P}+C_g^{AP}$ and $E_C^{AP}/E_C^{P}=C_T^{P}/C_T^{AP}\approx1.004$ for CrI$_3$. This affects only the peak widths, at the sub-percent level.

\textit{Temperature.}---We specify temperature by $k_BT/E_C^{P}$ ($0.05$ unless stated otherwise), and evaluate the AP state at the same physical temperature, $k_BT/E_C^{AP}=(k_BT/E_C^{P})\,E_C^{P}/E_C^{AP}$. Temperature sets how sharply the P and AP peaks are resolved, but not their positions or $V^*$.

\textit{Numerical evaluation.}---For each $n_g$, Eq.~(\ref{eq:G}) is summed over the charge states within $\pm3$ of the sweep range; states further away carry Boltzmann weights below $e^{-9E_C/k_BT}$. The Boltzmann factors are computed relative to the minimum of $U(N)$, and $F(x)$ is evaluated as $x/[-\mathrm{expm1}(-x)]$, with $F(0)=1$, for numerical stability. The mean charge is $\langle N\rangle=\sum_N N\,P_{\rm eq}(N)$. The P and AP responses are computed on a common $V_g$ grid using $n_g^{P,AP}=C_g^{P,AP}V_g/e$.

\textit{Parameters.}---Table~\ref{tab:SET_param} lists the inputs. $\varepsilon_M^{P,AP}$ come from our first-principles calculations. The island size and $C_T$ are illustrative values within the range reported for metallic (Al) SETs~\cite{rubia2025}; they set the absolute voltage and temperature scales but not $C_g^{AP}/C_g^{P}$ or the line shapes.

\begin{table}[h]
\centering
\begin{tabular}{ll}
\toprule
Quantity & Value\\ \midrule
$\varepsilon_M^{P}$, $\varepsilon_M^{AP}$ & Tab.~\ref{tab:gaps_eps} \\
$d_M$, $d_{\rm hBN}$ [Fig.~\ref{fig:SET}(a),(b)] & 100~nm, 0 (multilayer-dominant limit) \\
$d_M$, $d_{\rm hBN}$ [Fig.~\ref{fig:SET}(c),(d)] & 10~nm, 5~nm \\
$\varepsilon_{{\rm hBN},\perp}$ & 3.4 \\
Island area $A$ & $100\times100$~nm$^2$ \\
$C_T^{P}$ & 100~aF \\
$k_BT/E_C^{P}$ & 0.05 \\
$R_{l,r}$ (cotunneling estimate) & 0.1--1~M$\Omega$ \\ 
\bottomrule
\end{tabular}
\caption{\label{tab:SET_param}Parameters of the SET model.}
\end{table}

\textit{Assumptions.}---(i) Transport is sequential tunneling in linear response through junctions with $R_{l,r}\gg h/e^2$; cotunneling enters only through the estimate of the valley conductance. (ii) The magnetic configuration affects only $\varepsilon_M$; the source, drain and stray capacitances and the offset charge $q_0$ are identical in P and AP. For a symmetrically encapsulated multilayer, the magnetic-state-dependent charge redistribution in M carries no net dipole and leaves $q_0$ unchanged; a change in $q_0$ would appear as a rigid shift of the whole comb, distinguishable from the $N$-linear shift found here. (iii) The multilayer switches uniformly; with partial switching, $\varepsilon_M$ becomes an effective-medium average. (iv) The multilayer is insulating, so there is no free-carrier screening of the gate field. (v) The parallel-plate estimate of $C_g$ neglects fringing. For the parameters of Table~\ref{tab:SET_param} ($d_M$ comparable to the island size) fringing is not negligible and increases $C_g^P$ and $C_g^{AP}$ by a geometry-dependent factor of order a few. If the multilayer extends laterally beyond the island, the fringing field also traverses the multilayer, so $C_g^{AP}/C_g^P$, the line shapes and $G_P/G_{AP}$ are essentially unchanged, while $V^*$ and the peak spacing $e/C_g$ are reduced by the same factor. The quoted $V^*$ values are therefore order-of-magnitude (upper) estimates for the given stack. (vi) The island is a normal metal with a large density of states at $E_F$, so its quantum capacitance is effectively infinite and does not enter $C_g$. (vii) $\Delta E\ll k_BT$: for a $100\times100\times20$~nm$^3$ Al island, $\Delta E\approx0.2~\mu$eV, compared with $k_BT\approx40~\mu$eV at $0.47$~K.

\clearpage

\clearpage
\setcounter{figure}{0}\setcounter{table}{0}\setcounter{equation}{0}
\renewcommand{\thefigure}{S\arabic{figure}}
\renewcommand{\thetable}{S\Roman{table}}
\renewcommand{\theequation}{S\arabic{equation}}
\begin{center}
{\large\bfseries Supplementary Material: Spin-valve effect through electrostatic coupling}
\end{center}

\section*{Details on DFT calculations}
DFT results were obtained using the {\sc QUANTUM ESPRESSO} package \cite{QE-2017, QE-2009, doi:10.1063/5.0005082}. We performed GGA-PBE \cite{PBE1996} calculations, using norm-conserving pseudopotentials \cite{van_Setten_2018, Hamann2013} %\redmark{Check: cite only the libraries actually used. GBRV (garrity2014pseudopotentials) is ultrasoft and the Delta-test paper (doi:10.1126/science.aad3000) is not a pseudopotential set, so both were removed here.}
for structural relaxations, electronic structure calculations (such as band structures) and work functions. Calculations are collinear, without spin-orbit coupling, and without a Hubbard correction due to our interest in the differences between the P and AP configurations rather than in absolute band gaps. %, and the selection-rule argument of the main text is formulated in the absence of spin-orbit coupling. 
The plane-wave energy cutoffs were converged in the bilayer structures: for the CrX$_3$ (X = Cl, Br, I) bilayers, with AB stacking, to 80 Ry for the wave functions and 320 Ry for the charge density; and for NiBr$_2$ to 100 Ry for the wave functions and 400 Ry for the charge density. The same parameters were used for the heterostructures.% For the bader charge analysis results in the specific case of the CrX$_3$ heterostructures, we used PAW pseudopotentials \cite{Jollet2014}, to compute the all-electron charge densities, with values converged in the bilayer structures for the plane-wave energy cutoff of 50 Ry for the wave functions and 600 Ry for the charge density. The same parameters were chosen to perform the calculation of the heterostructures.

\begin{figure}
    \centering
    \includegraphics[width=\linewidth]{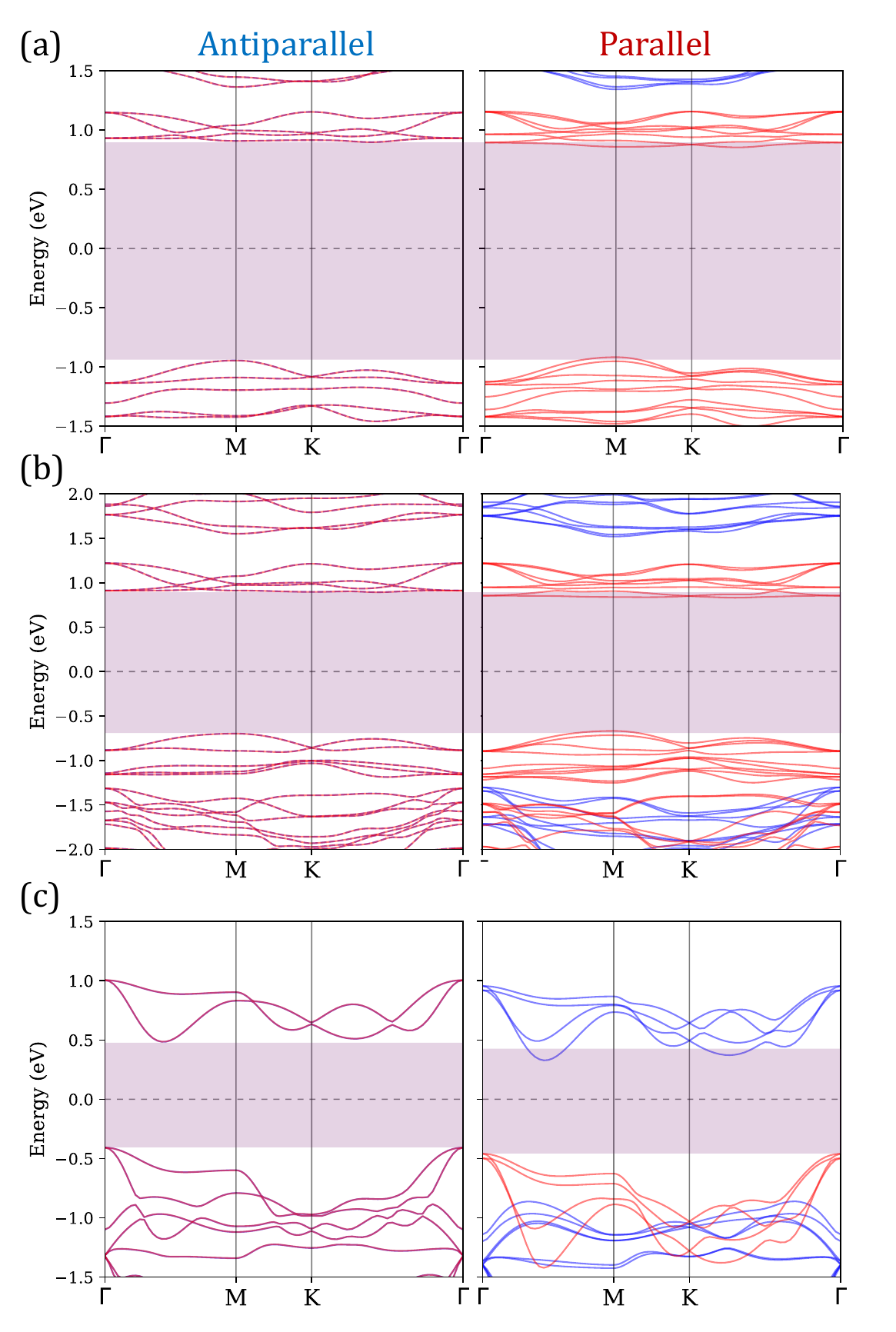}
    \caption{DFT band structures of insulating bilayers for antiparallel (left) and parallel (right) magnetic order of the spins between layers: (a) CrCl$_3$, (b) CrBr$_3$ and (c) NiBr$_2$. Blue (red) lines represent the minority (majority) spin channel. The shaded area represents the band gap of the antiparallel configuration, showing by comparison that the gap of the parallel configuration is smaller.}
    \label{fig:sup_band_gap}
\end{figure}

In-plane lattice parameters and internal atomic positions were relaxed for every individual system, i.e., the NiBr$_2$ and CrX$_3$ bilayers and graphene, in the magnetic ground state.
The P and AP configurations were obtained by setting parallel or opposite starting magnetizations on the magnetic atoms of the two layers, using the same geometry for both. All calculations (including heterostructures) include van der Waals corrections through the semiempirical Grimme DFT-D3 method \cite{Grimme2010}. Every isolated bilayer was treated as a metal (even though it is an insulator), using Gaussian smearing for the occupations, so that the same parameters could be used for the heterostructures, which are metallic. The bilayer and heterostructure cells include a converged vacuum of 15~\AA\ along $z$ to avoid interactions between periodic images. Lattice parameters for the different bilayers are given in Table~\ref{tab:latpar}. The Brillouin zone integrals for the relaxations were performed on converged 11$\times$11$\times$1 $k$-point grids for the CrX$_3$ bilayers and 12$\times$12$\times$1 for NiBr$_2$, with a Gaussian smearing of 0.001 Ry. The band structures of these systems were calculated with the same $k$-meshes as the relaxations, leading to the results in Fig.~\ref{fig:sup_band_gap}. For the isolated bilayers, the Fermi level lies at midgap, which we adopt as the reference for the work function, $\Phi=E_{\rm vac}-E_F$.

Bulk calculations, used for the band gaps and dielectric constants in Table~I of the main text, were performed for the AB stacking structure of each compound with 11$\times$11$\times$6 and 11$\times$11$\times$3 $k$-meshes for the Cr halides and NiBr$_2$ respectively, and the same cutoffs as above, in both the P and AP configurations (opposite starting magnetizations on alternate layers). The dielectric tensor was obtained from density-functional perturbation theory as implemented in {\sc QUANTUM ESPRESSO} for the bulk structures, with fixed occupations, at $\Gamma$; we report the $zz$ component of the electronic (clamped-ion) dielectric tensor $\varepsilon^\infty$. This relates to the optical gap $E_g^O$ in Table~I of the main text, which is the lowest energy separation between occupied and empty states of the same spin, i.e., the threshold for spin-conserving interband transitions; it coincides with the fundamental gap $E_g^F$ whenever both band edges have the same spin character.

\begin{table}[t]
    \centering
    \caption{Lattice parameters for the different isolated bilayers and graphene.}
    \label{tab:latpar}
    \begin{tabular}{cc}
        \toprule
        Material & Lattice Parameter (\AA) \\
        \midrule
        NiBr$_2$ & $3.67$  \\
        CrCl$_3$ & $5.98$ \\
        CrBr$_3$ & $6.35$ \\
        CrI$_3$  & $6.88$  \\
        Graphene & $2.47$ \\
        \bottomrule
    \end{tabular}
\end{table}

The heterostructures were constructed using commensurate supercells (SCs), with the lattice mismatch accommodated by applying equal and opposite strain to the bilayers and graphene. The resulting supercells and strain values are given in Table~\ref{tab:supercells}; the graphene--bilayer distances $d$ were relaxed. Strain values are reported with respect to the bilayer: negative (positive) values correspond to compressive (tensile) strain in the bilayer, while graphene experiences an equal tensile (compressive) strain. The P and AP configurations of each heterostructure are computed at the same strain and geometry, so that their difference is not affected by it.

\begin{table}[t]
    \centering
    \caption{Lattice mismatch (strain) and commensurate supercells (SC) for the graphene/bilayer heterostructures.}
    \label{tab:supercells}
    \begin{tabular}{cccc}
        \toprule
        Bilayer & Strain (\%) & Bilayer SC & Graphene SC \\
        \midrule
        NiBr$_2$ & $0.41$ & $2\times2$ & $3\times3$ \\
        CrCl$_3$ & $-2.45$ & $\sqrt{3}\times\sqrt{3}$ & $4\times4$ \\
        CrBr$_3$ & $-1.45$ & $2\times2$ & $5\times5$ \\
        CrI$_3$  & $1.68$ & $\sqrt{3}\times\sqrt{3}$ & $5\times5$ \\
        \bottomrule
    \end{tabular}
\end{table}

\begin{figure}
    \centering
    \includegraphics[width=\linewidth]{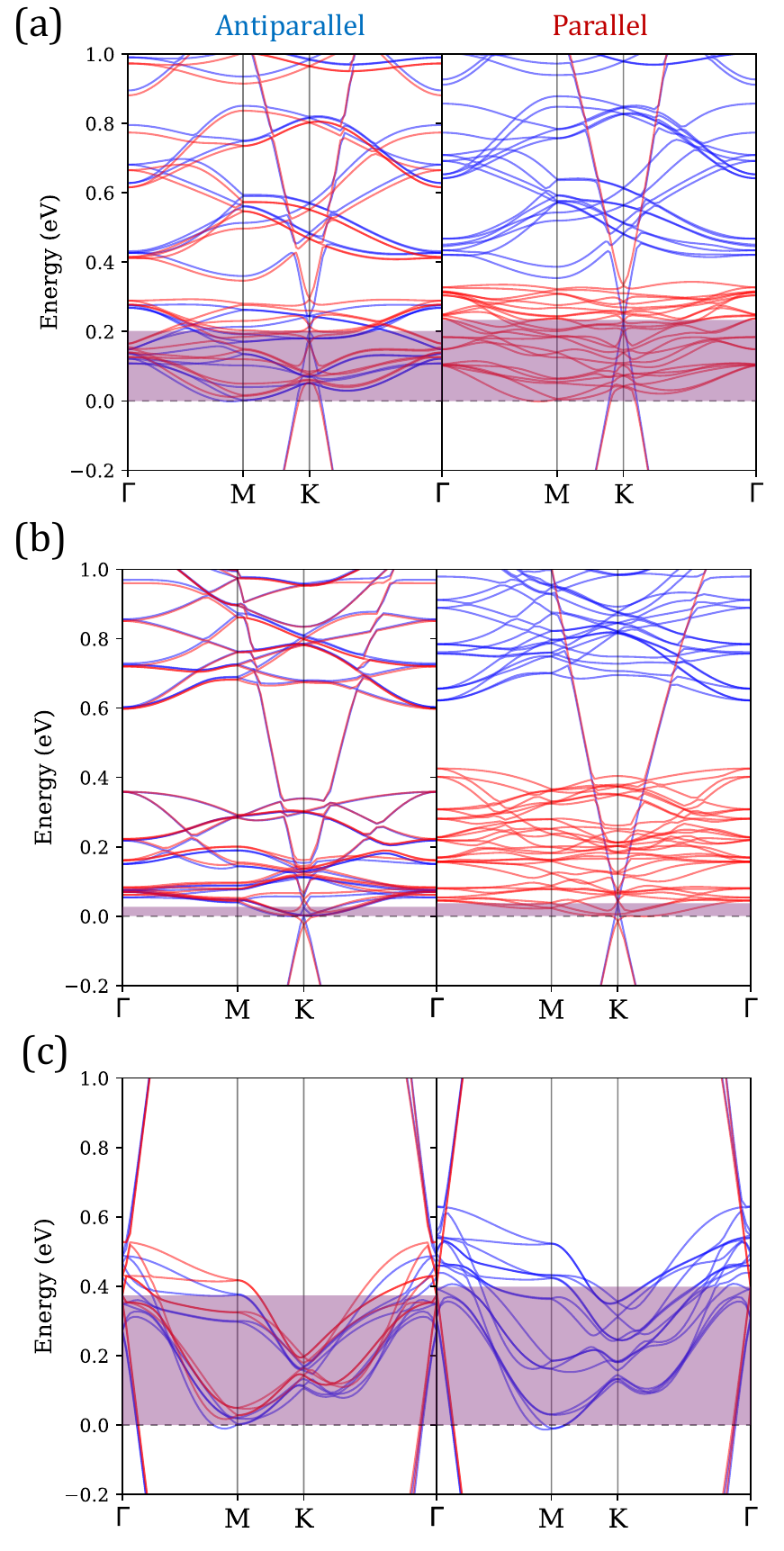}
    \caption{Band structures of heterostructures consisting of graphene on top of a magnetic insulating bilayer: (a) CrCl$_3$, (b) CrBr$_3$ and (c) NiBr$_2$. Blue (red) lines represent the minority (majority) spin channel. Left (right) panels correspond to antiparallel (parallel) order of the spins between the layers of the insulating bilayer. The shaded area is a guide to the eye showing where the graphene Dirac cone falls within the bands of the insulating bilayer.}
    \label{fig:sup_hetero_struct}
\end{figure}

For the heterostructure calculations, the structural parameters were those obtained from the relaxation in the magnetic ground state. We used Gaussian smearing of 0.001 Ry in all systems. The Fermi energy was determined from a non-self-consistent calculation with the tetrahedron method \cite{Blochl1994}. The band structures were calculated on 6$\times$6$\times$1 $k$-meshes for CrBr$_3$, 7$\times$7$\times$1 for CrCl$_3$ and CrI$_3$, and 12$\times$12$\times$1 for NiBr$_2$, leading to Fig.~\ref{fig:sup_hetero_struct}. For the work functions, all calculations included dipole corrections \cite{bengtsson1999dipole, neugebauer1992adsorbate}, and the $k$-meshes were increased to 13$\times$13$\times$1 for the CrX$_3$ systems and kept at 12$\times$12$\times$1 for NiBr$_2$. Work functions were obtained from the planar-averaged electrostatic potential along $z$, with $E_{\rm vac}$ taken from its plateau in the vacuum region; for the heterostructures, $\Phi_{graphene}$ and $\Phi_{bilayer}$ correspond to the plateaus on the graphene and bilayer sides, which share a common Fermi level.

Table~\ref{tab:WF_hetero} lists the work-function differences $\Delta\Phi=\Phi_{\rm bilayer}-\Phi_{\rm graphene}$ obtained in this way for the AP and P configurations of each heterostructure, together with the resulting graphene conductivity ratio of Eq.~(3) of the main text. Since both configurations share the same geometry, the graphene--bilayer distance $d$ in Eq.~(2) of the main text cancels in the ratio.

\begin{table}[t]
    \centering
    \caption{Work-function difference $\Delta\Phi=\Phi_{\rm bilayer}-\Phi_{\rm graphene}$ across the graphene/bilayer heterostructures in the AP and P configurations, and the resulting graphene conductivity ratio $\sigma_P/\sigma_{AP}=\Delta\Phi_P/\Delta\Phi_{AP}$ (Eq.~(3) of the main text).}
    \label{tab:WF_hetero}
    \begin{tabular}{cccc}
        \toprule
        Bilayer & $\Delta\Phi_{AP}$ (meV) & $\Delta\Phi_{P}$ (meV) & $\sigma_P/\sigma_{AP}$ \\
        \midrule
        CrCl$_3$ & 188.1 & 225.6 & 1.200 \\
        CrBr$_3$ & 193.3 & 231.7 & 1.199 \\
        CrI$_3$  & 322.8 & 385.3 & 1.194 \\
        NiBr$_2$ & 625.3 & 713.9 & 1.142 \\
        \bottomrule
    \end{tabular}
\end{table}

\section*{Electronic structure discussion}
The complete spin polarization of the band edges in the Cr trihalides can be traced directly to the atomic physics of the Cr$^{3+}$ ion. Chromium enters CrX$_3$ in a $+3$ oxidation state (each halide X$^-$ carrying charge $-1$), giving a $3d^3$ configuration. Cr$^{3+}$ sits at the center of an (edge-sharing) CrX$_6$ octahedron, so the five $d$ orbitals are split by the octahedral crystal field into a lower $t_{2g}$ triplet and a higher $e_g$ doublet. Hund's rule favors the high-spin configuration in which all three $d$ electrons singly occupy the $t_{2g}$ orbitals with parallel spin, %$t_{2g}^{3}e_g^{0}$,
giving each Cr a local moment $S=3/2$ ($\sim 3\,\mu_B$). The empty $e_g$ level is itself spin-split by the same Hund's exchange. Because the exchange splitting is comparable to, or larger than, the $t_{2g}$--$e_g$ crystal field splitting, the empty \emph{majority}-spin $e_g$ level drops below the empty \emph{minority}-spin $t_{2g}$ level. %The conduction band of Cr trihalides is therefore formed exclusively by these majority-spin Cr $e_g$ states -- a band with essentially pure, $100\%$ spin-polarized Cr character.

The valence band maximum is formed by the filled majority-spin $t_{2g}$ states. Because these occupied $t_{2g}$ levels sit close in energy to the occupied halide $p$ states, the two hybridize substantially, giving the valence band mixed Cr/halide character (typical charge-transfer-insulator behavior), whereas the empty $e_g$ states are pushed well above the halide $p$ manifold and hybridize only weakly with the ligands. %The valence band is thus also $100\%$ spin polarized, but unlike the conduction band -- with appreciable halide weight.

In the P state, interlayer hopping hybridizes and broadens both band edges. Their relative contribution to the gap reduction depends on the orbital character at the interlayer contact: the two layers are held together only by van der Waals forces, and the interlayer gap is bridged by the outward-facing halide atoms, while the Cr atoms, sandwiched inside each CrX$_6$ layer, are comparatively shielded from it. The valence band, with its substantial halide-$p$ weight, and the conduction band, of predominantly Cr $e_g$ character, therefore hybridize to different extents, and the gap change will in general involve an asymmetric shift of the two band edges. 

NiBr$_2$ behaves differently. Ni$^{2+}$ has a $3d^8$ configuration, with fully occupied $t_{2g}$ levels and two electrons in the majority-spin $e_g$ level ($S=1$). The lowest empty states are therefore the minority-spin $e_g$ orbitals, so that the conduction band minimum has minority-spin character while the valence band maximum is majority-spin. In the P state the fundamental gap is thus a spin-flip transition, which does not contribute to $\varepsilon$ in the absence of spin-orbit coupling. In the AP state, the minority-spin conduction band of one layer has the same spin as the majority-spin valence band of the other, so the lowest transition becomes spin-allowed, consistent with $E_g^F=E_g^O$ in the AP state (Table~I of the main text). This transition is interlayer and polarized along $z$, so it contributes to $\varepsilon^{\infty}_{zz}$ through the interlayer hybridization of the band edges. Consistent with this, the enhancement of $\varepsilon^\infty$ from P to AP is predominantly out of plane: $\varepsilon^\infty_{zz}-1$ increases by 31\%, compared with 8\% for $\varepsilon^\infty_{xx}-1$.

\section*{Work function and band gap}

A relationship between the work function and the electronic band gap of an intrinsic insulator follows from the definition of the work function,
\begin{equation}
    \Phi = E_{\rm vac} - E_F = (E_{\rm vac}-eV_{\rm bulk})+(eV_{\rm bulk}-E_F),
    \label{work-function}
\end{equation}
whose change between the two magnetic configurations, $\delta X\equiv X^P-X^{AP}$, is
\begin{equation}
    \delta \Phi = \delta(E_{\rm vac}-eV_{\rm bulk})+\delta(eV_{\rm bulk}-E_F).
    \label{eq:dPhi}
\end{equation}
The first term corresponds to the surface dipole,
\begin{equation}
    E_{\rm vac}-eV_{\rm bulk}=e\Delta V_{\rm surf}=-\frac{e\,\sigma_{dip}d_{dip}}{\varepsilon_0\varepsilon},
    \label{eq:surf}
\end{equation}
where $\sigma_{dip}$ is the areal charge density of the surface dipole layer and $d_{dip}$ its thickness, and the second to the bulk electronic structure,
\begin{equation}
    eV_{\rm bulk}-E_F = eV_{\rm bulk}-E_v-\frac{E_g}{2},
    \label{eq:bulk}
\end{equation}
where we have used $E_F = (E_c+E_v)/2=E_v+E_g/2$, with $E_v$ ($E_c$) the valence band maximum (conduction band minimum).

To see how the gap change enters, let a fraction $f$ of it come from the conduction band and the remainder from the valence band, $\delta E_c=f\,\delta E_g$ and $\delta E_v=-(1-f)\,\delta E_g$, measured relative to $eV_{\rm bulk}$. Then $\delta(eV_{\rm bulk}-E_v)=(1-f)\,\delta E_g$, and Eq.~(\ref{eq:dPhi}) becomes
\begin{equation}
    \delta \Phi \approx e\delta\Delta V_{\rm surf}+\left(\tfrac{1}{2}-f\right)\delta E_g .
    \label{eq:dPhi_f}
\end{equation}
This splits $\delta \Phi$ into a screening-dependent surface term and a hybridization-driven band term, whose weight depends on how asymmetrically the band edges shift. A symmetric opening or closing of the gap ($f=1/2$) leaves only the surface term, $\delta\Phi\approx e\delta\Delta V_{\rm surf}$, whereas a gap change carried entirely by the conduction (valence) band gives $\delta\Phi\approx e\delta\Delta V_{\rm surf}\mp\delta E_g/2$. In general, therefore, a magnetic-state-dependent gap produces a magnetic-state-dependent work function, with a sign and magnitude set by the band-edge asymmetry and by the accompanying change in the surface dipole.

For all four materials, $|\delta\Phi|$ (Table~I of the main text) is considerably smaller than $|\delta E_g|/2$, indicating that, unless the band edges shift almost symmetrically, the surface-dipole term compensates a significant part of the band term.

\end{document}